# Spectropolarimetry as a Means to Address Cloud Composition and Habitability for a Cloudy Exoplanetary Atmosphere in the Habitable Zone

Robert A. West[1,2], Philip Dumont[3], Renyu Hu[3,4], Vijay Natraj[3], James Breckinridge[4,5,6], Pin Chen[3]

## ABSTRACT

In our solar system, the densely cloud-covered atmosphere of Venus stands out as an example of how polarimetry can be used to gain information on cloud composition and particle mean radius. With current interest running high on discovering and characterizing extrasolar planets in the habitable zone where water exists in the liquid state, making use of spectropolarimetric measurements of directly-imaged exoplanets could provide key information unobtainable through other means. In principle, spectropolarimetric measurements can determine if acidity causes water activities in the clouds to be too low for life. To this end, we show that a spectropolarimeter measurement over the range 400 nm – 1000 nm would need to resolve linear polarization to a precision of about 1% or better for reflected starlight from an optically thick cloud-enshrouded exoplanet. We assess the likelihood of achieving this goal by simulating measurements from a notional spectropolarimeter as part of a starshade configuration for a large space telescope (a HabEx design, but for a 6 m diameter primary mirror). Our simulations include consideration of noise from a variety of sources. We provide guidance on limits that would need to be levied on instrumental polarization to address the science issues we discuss. For photon-limited noise, integration times would need to be of order one hour for a large radius (10 Earth radii) planet to more than 100 hours for smaller exoplanets depending on the star-planet separation, planet radius, phase angle and desired uncertainty. We discuss implications for surface chemistry and habitability.

## 1. INTRODUCTION

Interest in exoplanets has exploded in recent years, especially after the successful prime and extended Kepler missions. The inventory of published papers on exoplanets has also increased exponentially. Perhaps the primary driver for this is the possibility that life may exist on planets in the habitable zone (by definition, the configuration of host star radiant spectral flux and planet chemistry and temperature conducive to life as we know it). The recent ASTRO2020 "Pathways to Discovery in Astronomy and Astrophysics for the 2020s" by the National Academies Press (hereafter the Astro2020 document) provides a compelling rationale for future investment in exoplanet research in general, and for direct imaging of exoplanets in particular.

---

[1] Corresponding author robert.a.west@jpl.caltech.edu
[2] 8550 N Maple Ave, Fresno, CA 93720
[3] Jet Propulsion Laboratory, California Institute of Technology, Pasadena, CA 91109
[4] California Institute of Technology, Pasadena, CA 91125
[5] The University of Arizona, Tucson, AZ 85717
[6] Deceased

Direct imaging of exoplanets offers some desirable attributes not available from transit observations used to characterize planetary atmospheres via spectroscopy. Photons scattered from a planetary atmosphere or surface observed from direct imaging sense the atmosphere at deeper levels than do photons traversing the atmosphere in transit geometry. Direct imaging also provides the possibilities of detecting surface mineralogy, (Hu et al. (2012), for a planet with an optically thin or no atmosphere) and bodies of liquid on the surface, via specular reflection (Robinson et al. 2010). The target list for direct imaging expands the option space by not requiring the restrictive geometry of a transit. For planets orbiting of order 1 AU from their host star, transits are hard to come by, whereas the angular separation required for direct image favors planets that are not too close to their host star. To be in the habitable zone means that planet detection/characterization for M star hosts are favored by transits, whereas for a G star like our sun they are favored by direct imaging, provided that the technology can sufficiently suppress light from the host star.

Transit spectroscopy from the near-ultraviolet to the near-infrared has provided information on the composition of exoplanetary atmospheres, but with limitations and degeneracies in some derived parameters (Seager 2008; Fu et al. 2018). Information on exoplanets can also come from other types of observations, including eclipses observed at thermal-infrared wavelengths, radial velocity measurements, microlensing, and precision astrometry. These techniques also give information about atmospheric composition, temperature and longitudinal temperature gradients, planet mass, radius, and orbit.

Here we explore what polarization measurements can tell us for a very specific case: a Venus-like exoplanet with global cloud cover. Exoplanet Kepler 1649b may be one example ($R_p/R_E$ = 1.08 +- 0.15; orbital semi-major axis = 0.0514 +- 0.0028 au, Angelo et al. 2017). Another might be Kepler 69c which orbits a G4V-type star every 242.5 days ($R_p/R_e$ = 1.7 +0.34 −0.23; orbital semi-major axis = 0.64 +0.15 −0.11 au, Barclay et al. 2013). Exo-Venus examples are a class of habitable-zone planets whose polarization signature can bear on the potential for life on these planets. As part of this investigation, we also look at what signal/noise ratio would be required, and what would be feasible for a notional 6-m space telescope based on the 4-m HabEx concept (Gaudi et al. 2020), and we comment on how instrumental polarization for such a mission would need to be controlled.

Observations of the polarization state of exoplanets potentially offer unique information. Karalidi et al. (2011) and (2012) show how water clouds on an earth-like exoplanets can be detected using polarization measurements at phase angles (supplementary to the scattering angles) in and near the rainbow feature in the phase angle range 30° – 40°. Polarization might also be used simply to detect exoplanets, by virtue of near-zero polarization of the host star, in the ideal case. Seager et al. (2000) explored this approach for close-in hot Jupiters, for idealized atmospheric polarization cases. An atmosphere dominated (optically) by Rayleigh scattering by gas molecules is a best-case idealization because Rayleigh scattering is almost fully polarized (for single scattering) at a scattering angle of 90°. In practice, the polarization signal will be diminished due to either (1) multiple scattering for an optically thick atmosphere, or (2) low signal/noise for an optically thin atmosphere or for a thick atmosphere with significant absorption from gas or absorbing particles mixed with the gas (e.g., Madhusudhan & Burrows 2012). For a conservatively scattering optically thick Rayleigh atmosphere, the disk-integrated maximum polarization is close to 30% (Horak 1950). Wiktorowicz & Stam (2015) gave a comprehensive review of observational searches (up to 2015, using polarimetery) for close-in exoplanets. The case of HD 189733 has a long and controversial history (Bott et al. 2016).

## 2. VENUS-LIKE CLOUD COVER, AND HABITABILITY

Two features of a Venusian atmosphere distinguish it from other solar system planets. First, by virtue of having clouds composed of liquid droplets, and temperature (in the clouds) conducive to liquid water, exoplanet analogous to Venus are by definition potentially habitable, and therefore of special interest as targets to search for direct imaging.

Second, Venus has a global cloud cover of spherical liquid drops. That feature provides a means, via Mie theory for light scattering by spheres, to sense cloud particle refractive index and mean radius (Hansen & Hovenier 1974). While water cloud properties in the Earth's atmosphere also can be retrieved in the same way, integrated whole-disk reflectance, appropriate for directly-imaged exoplanets, contains contributions from surface land, water, ice, and atmospheric dust and ice particles. These other contributions cannot be addressed using Mie theory. However, Karalidi et al. (2011) and (2012) showed how polarization can be used to infer liquid water clouds for an earth-like exoplanet. Spectrophotometry can be also used to tease out information on them, given sufficient integration time (Tinetti et al. 2006; Gu et al. 2021). Various aspects for using polarization to retrieve surface water and atmospheric components for Earth-like exoplanets has also been addressed by others (Stam 2008; Karalidi & Stam 2012; Sterzik et al. 2012; Wiktorowicz & Stam, 2015; García Muñoz 2018; Sterzik et al. 2019; Trees & Stam, 2019).

In our solar system, the only other planets with atmospheres whose particle properties can be derived from polarimetry are Titan and the polar hazes of Jupiter and Saturn; for those we use scattering codes relevant to fractal aggregates of monomers in the Rayleigh regime (West & Smith 1991; Tomasko et al. 2009). Comprehensive reviews of polarization studies for other planets in our solar system are provided by Kaydash et al. 2015, and by West et al. 2015.

In the wavelength range 0.4 - 1 µm small droplets of liquid water and sulfuric acid are spectrally smooth and indistinguishable. In the early 1970s polarization measurements provided a breakthrough in understanding the composition and mean particle radius of the Venusian cloud tops (Hansen and Hovenier 1974). Hansen and Hovenier (1974) used polarization measurements at several wavelengths and over a wide range of phase angles to determine both the real part of the refractive index (1.46 +- 0.015 at 0.365 µm to 1.43 +- 0.015 at 0.99 µm) and mean radius (1.05 +- 0.10 µm at a mean atmospheric pressure of 50 mb). The retrieved refractive indices are consistent with concentrated sulfuric acid, and inconsistent with other candidate compositions proposed at the time of publication. The focus of this paper is to assess if a similar study could be carried out for exoplanets, and if so, could such an effort be relevant to the question of habitability in the clouds of an exoplanet.

By virtue of its sensitivity to acid concentration, knowledge of the polarizing properties of the Venus clouds leads to a conclusion that life as we know it cannot exist in the clouds of Venus. Figure 1 from Hallsworth et al. (2021) shows the limits of known life with respect to sulfuric acid concentration and water activity. Water activity is equivalent to the equilibrium vapor pressure of an aqueous sample normalized by that of pure water. Water activity is directly related to dissolved ion concentrations, or salinity, and it is a measure of the availability of water for life in an acidic environment. In the above-mentioned figure, limits of life as we know it are shown as two points on the plot, one for low temperature, and one for high temperature. In the low temperature (~25 C) case, life can exist with sulfuric acid concentration no higher than about 40%. For the high-T (60 C) case, the limit is 10%. Hallsworth et al. (2021) illustrated that, due to concentrated sulfuric acid, water activity in the Venusian cloud environment is far below the lower limit that can support known life.

## 3. EXOPLANET DIRECT IMAGING AND POLARIMETRY

Anticipating that direct imaging and polarimetry of exoplanets in the habitable zone may someday be possible, we examined expected polarization for a Venus-like exoplanet having global, optically thick, cloud cover, as a function of sulfuric acid concentration up to 50%. That covers the range within which life on earth is possible. With these calculations we address the question of how precisely polarization needs to be measured to distinguish the different cases.

Because liquid droplets are spheres, neglecting deformation from perturbations, we use a Mie code (Hansen & Travis 1974) to calculate scattering matrix elements for particles with refractive indices corresponding to acid concentrations 0%, 25% and 50%. These were obtained from online files from the University of Oxford Earth Observation Data Group http://eodg.atm.ox.ac.uk/ARIA/search. For pure water we used the file named H2O_Stegelstein_1981.ri (Segelstein, D. 1981). For a 25% sulfuric acid concentration we used the file H2SO4_25_Palmer_1975_R.ri (Palmer & Williams 1975). Values for a 50% weight concentration of $H_2SO_4$ came from the file H2SO4_50_Palmer_1975_R.ri (Palmer & Williams 1975).

We computed phase matrix elements on a four-dimensional grid (phase angle or scattering angle 0 - 180° in steps of 1°, wavelength from 0.4 to 1.8 μm in steps of 0.1 μm, particle mean radius from 0.25 μm to 4 μm in steps of factors of two, and for the three compositional cases. Particles have a polydisperse (Gamma) distribution with a variance of 0.07 as retrieved for Venus by Hansen and Hovenier, 1974. The behavior of the polarization as a function of particle size, acid concentration, and phase angle can be gauged from Figure 1. Each panel in Figure 1 shows the phase angle behavior of the polarization (for single scattering for a single particle in the left column and for total (single plus multiple) scattering integrated over the disk in the right column) for three compositions: water (solid curve), 25% sulfuring acid (dash-dot) and 50% sulfuring acid (long dashes). The calculations are for wavelength 0.7 μm.

Polarization from single scattering for the smallest particles considered here (particle mean radius <r> = 0.25μm) have a broad positive polarization peak (the electric vector oscillates in the plane perpendicular to the scattering plane) reaching a maximum around 75° - 80°. Smaller particles (or longer wavelengths) produce polarization that tends toward the Rayleigh limit, peaking at or near 100% at phase 90°. When total scattering from a spherical planet is considered (right column), the peak polarization shifts to higher phase angles as the ratio of single/total scattering increases at the higher phase angles. Multiple scattering scrambles polarization and thereby reduces it (note the scale difference between the left and right columns of Figure 1).

As particle mean radius increases a rainbow feature becomes prominent at phase angles in the range 30° - 40°, reaching 80% polarization for (single scattering) for sulfuric acid. A moderately strong negative branch also appears at lower phase angles. A secondary polarization peak corresponding to the secondary rainbow can be seen near 60° phase for the larger particles. The separation of these two peaks is greatest for the 50% sulfuric acid case, due mostly to the primary rainbow shift to smaller phase angles as the acid concentration increases.

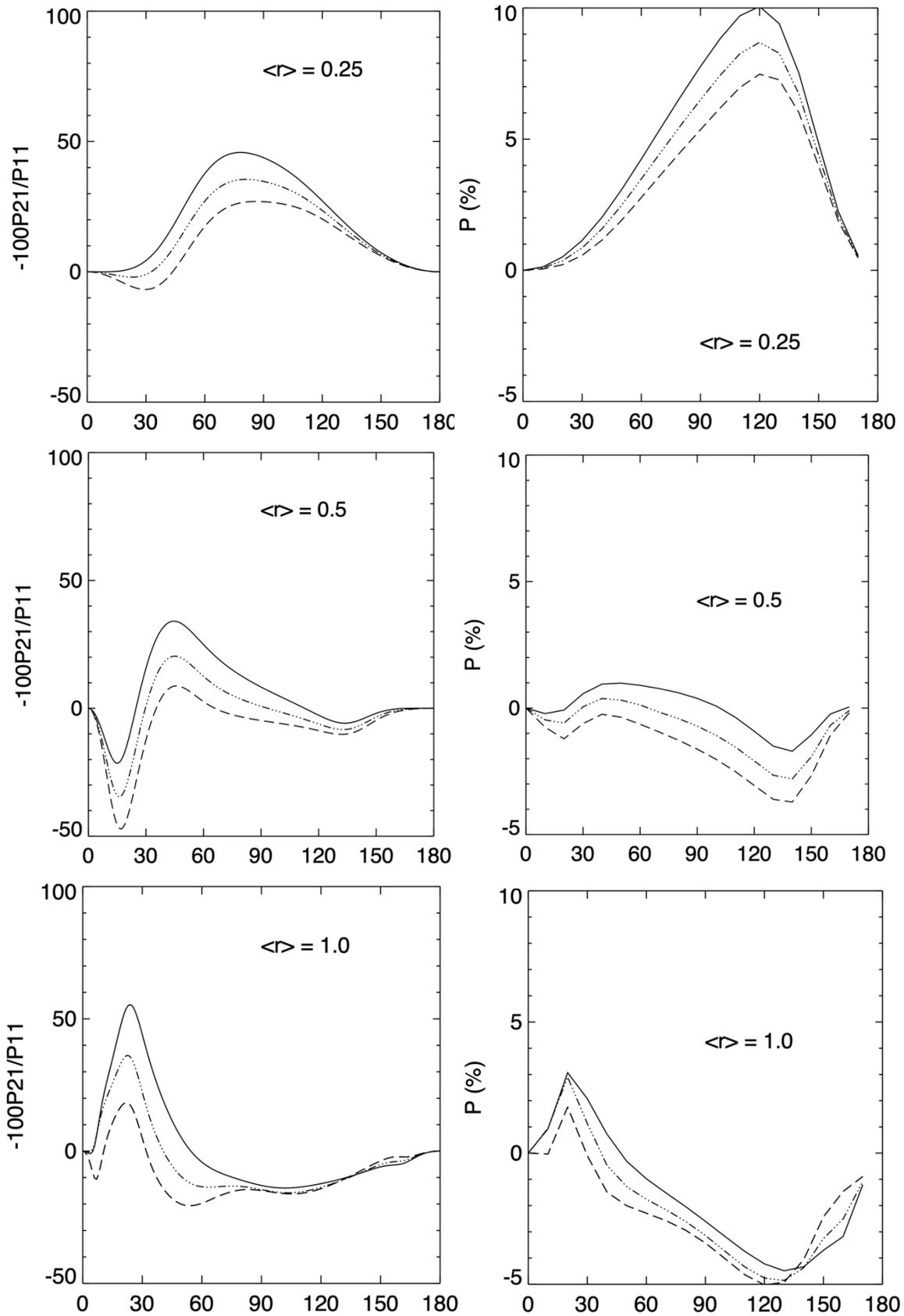

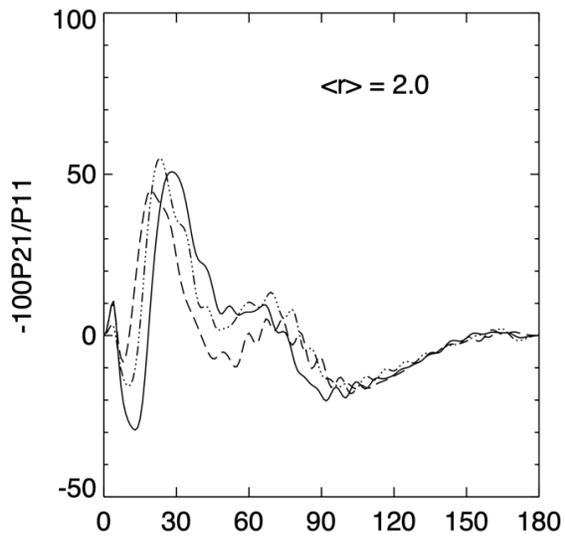
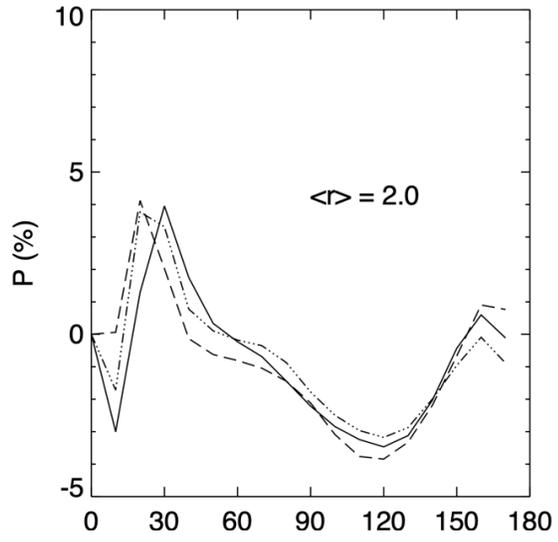
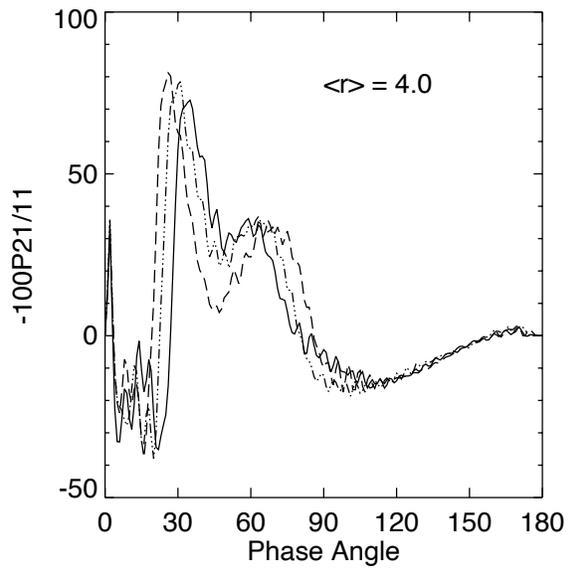
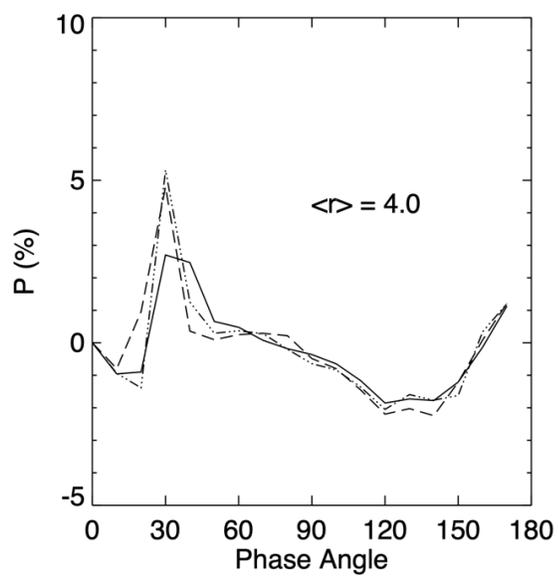

***Figure 1*** *Left column: linear polarization (phase matrix elements -P21/P11) for single scattering for three compositions and five particle mean radii <r>. All results are for wavelength 0.7 µm. Results are shown for steps of one degree in phase angle. Right column: corresponding net polarization, for single plus multiple scattering, and disk-averaged for the whole planet. Results are shown for ten-degree steps in phase angle. The solid curve is for water, the dash-dot for 25% $H_2SO_4$, and the long dashes for 50% $H_2SO_4$.*

We used the VLIDORT code (Spurr 2006) for multiple scattering calculations for an optically thick cloud (cloud optical depth >> 1, as for Venus). The code uses the discrete ordinates method to calculate Stokes parameters I, Q, and U for a specified scattering geometry (a combination of incidence angle, emission angle, and azimuthal angle). The degree of linear polarization is given by Sqrt($Q^2 + U^2$)/I. To obtain disk-integrated polarization these were computed on a set of points on the sphere and were summed according to Horak's cubature formulas (Horak 1950) for phase angles (phase angle equals 180° minus the scattering angle) from 0 to 170° in steps of 10°. The Stokes parameters calculated by VLIDORT are referenced to the meridional plane and need to be rotated to the scattering plane prior to summing. The formula for that rotation is given by Hansen and Travis 1974.

Multiple scattering contributes most of the light scattered by an optically thick, conservatively-scattering atmosphere, especially at low phase angles. In the 0.4 – 1 µm wavelength domain, water and sulfuric acid are nearly conservatively scattering and cannot be distinguished from spectra alone. At wavelengths longer than 1 µm both water and sulfuric acid have spectral absorption features. At all wavelengths, polarization features for optically thick atmospheres are suppressed by multiple scattering.

### 3.1 Particle Radius/Refractive Index Degeneracy and Retrieval Fidelity

For spherical liquid drops, particle radius and refractive index together determine polarization of the scattered light, so it is important to understand if their effects can be separated. More generally, it is important to determine how well acid concentration can be retrieved given the uncertainties of the measurements. Hansen and Hovenier (1974) were able to retrieve both properties of the Venusian cloud particles from precision polarization measurements over a range of phase angles and wavelengths. Here we assess how well this can be done for an exoplanet under a variety of possible observational configurations. We do this by simulating signals for one thousand measurement sets, where each measurement set is a set of simulated polarization values at N wavelengths and M phase angles. This provides a means for judging confidence in the retrieved acid concentration, and the best strategy for wavelength and phase angle coverage, which in turn map into instrument parameters, such as the telescope-starshade distance (one for the wavelength range 0.4 – 1.0 µm which we call the shortwave band, and another for the range 1.0 – 1.8 µm – the longwave band) and the number of telescope visits required for phase angle coverage. Within each wavelength band a set of measurements, with bandpass 100 nm, are simulated at 100-nm intervals which is sufficient to resolve polarization signatures that depend on wavelength.

Our selection of sample particle effective radii as shown in Figure 1 is somewhat arbitrary but is sufficient to illustrate the resulting uncertainties and spans, by a factor of 4 to lower and higher values, the value 1.05 µm derived for Venus by Hansen and Hovenier. In a planetary atmosphere particle mean radius is determined by the concentration of cloud condensation nuclei, among other factors, that are difficult or impossible to predict for exoplanets. Following Hansen and Hovenier, our computations use the Gamma size distribution with effective variance equal to 0.07. Hansen and Hovenier explored different values of the variance in the context of observations. Without observations such an exploration is not well justified because the effective radius dominates.

One thousand synthetic measurement sets were generated for each combination of particle effective radius and acid concentration (via the corresponding refractive index). Each measurement set consists of the polarization calculated for each combination of acid concentration and particle radius, plus Gaussian noise with polarization uncertainty σ. Noise samples were generated using a random number generator for a normal distribution and differ for each of the 1000 trials. From the collection of these synthetic measurement sets for each case we calculated a reduced χ2 for every combination of effective particle radius and acid concentration. Computations were made for five effective particle radii in steps of a factor of 2 from 0.25 µm to 4 µm (five values). These, combined with water and two acid concentration cases (25%, 50%), yield fifteen models. Reduced $\chi^2$ values were calculated for each of these measurement sets.

$$\chi^2 = \frac{1}{(N-2)} \sum_j \frac{(M_j - P_j)^2}{\sigma_j^2}$$

Where N is the number of synthetic measurements for a measurement set (number of wavelengths times number of phase angles), the value 2 accounts for the number of parameters in the fitting process (composition and particle effective radius), $M_j$ is the (synthetic) measurement for a particular wavelength and phase angle, and $P_j$ is the true polarization, a function of the model parameters, phase angle, and wavelength. Although in practice σ can depend on the wavelength and phase angle, for simplicity our simulations assume that each wavelength/phase angle has the same value.

The simulations were done for three values of σ: 1.0%, 0.5%, and 0.2% (relative to the total flux from the planet) to explore the range of uncertainty that would be required to differentiate among water, 25% acid, and 50% acid. What constitutes an acceptable level of confidence in the determination is subjective. Here we provide statistics for two metrics which might be used separately or together to retrieve acid concentration. The same could be done for particle effective radius but for habitability considerations acid concentration is the key parameter.

The first of these metrics is the probability that the model with lowest $\chi^2$ for a given measurement set is not the correct model, i.e. the model that was used to create the measurement set. This metric is relevant because a common retrieval method is to search for the model or set of parameters that minimizes $\chi^2$. But we also want to have some sense of how shallow such a minimum is. For that we have a second metric which is the probability that some model, not the correct one, has $\chi^2$ less than 2. It was chosen to allow for the fact that models with reduced $\chi^2$ less than 2 have a non-

| Table 1 Probabilities for Mis-identifying Acid Concentration | | | | | | |
|---|---|---|---|---|---|---|
| Simulated Observation Parameters | P1 | P2 | P1 | P2 | P1 | P2 |
| | $\sigma = 1\%$ | | $\sigma = 0.5\%$ | | $\sigma = 0.2\%$ | |
| SW  90. | 0.0419 | 0.1122 | 0.0122 | 0.0337 | 0.0013 | 0.0039 |
| LW  90. | 0.0080 | 0.0772 | 0.0004 | 0.0066 | 0.0000 | 0.0010 |
| SW+LW  90. | 0.0081 | 0.0766 | 0.0004 | 0.0064 | 0.0000 | 0.0010 |
| SW  60. | 0.0482 | 0.1354 | 0.0085 | 0.0248 | 0.0002 | 0.0012 |
| LW  60. | 0.0124 | 0.0963 | 0.0008 | 0.0118 | 0.0000 | 0.0010 |
| SW+LW  60. | 0.0117 | 0.0964 | 0.0008 | 0.0115 | 0.0000 | 0.0010 |
| SW  60. 90. 120. | 0.0106 | 0.1290 | 0.0012 | 0.0250 | 0.0000 | 0.0010 |
| LW  60. 90. 120. | 0.0008 | 0.0877 | 0.0000 | 0.0023 | 0.0000 | 0.0010 |
| SW+LW  60. 90. 120. | 0.0007 | 0.0872 | 0.0000 | 0.0023 | 0.0000 | 0.0010 |
| SW  30. 40. 50. | 0.0075 | 0.0976 | 0.0005 | 0.0162 | 0.0000 | 0.0010 |
| LW  30. 40. 50. | 0.0023 | 0.0981 | 0.0001 | 0.0161 | 0.0000 | 0.0010 |
| SW+LW  30. 40. 50. | 0.0024 | 0.0979 | 0.0001 | 0.0164 | 0.0000 | 0.0010 |
| Notes for Table 1. P1 is the probability that the retrieved acid concentration does not match the true acid concentration for retrievals that use minimum $\chi 2$ as the criterion. P2 is the probability that $\chi 2$ is less than 2 for the cases having acid concentration not equal to the true acid concentration. The parameter $\sigma$ is the standard deviation (uncertainty) for synthetic polarization measurements. In the column for simulated observation parameters SW = Short Wave (0.4-1.0 µm), LW = Long Wave (1.0-1.8 µm), and the numbers show the single or multiple phase angles (degrees) combined for the simulated observation. | | | | | | |

negligible probability of being the correct ones and could reasonably be categorized as possible solutions. As $\chi^2$ increases, a model becomes less and less likely to be the correct one. In practice, we use these metrics in a statistical sense to make a judgement on what precision would be needed for an acceptable determination of acid concentration.

If the measurement uncertainty is 1%, column 2 of Table 1 shows probability P1 in the range of 0.07% to 5% depending on which observational configuration is chosen. The P2 probability is in the range 8% to 14%. Lower probabilities are obtained if measurement uncertainties are reduced as shown in Table 1. Probabilities for longwave measurements are smaller than those for shortwave measurements due to greater spectral structure in the longwave set. Combining longwave and shortwave measurements does not significantly improve (lower) the probabilities relative to longwave only measurements. In the next section we explore how measurement uncertainties in shortwave measurements map to required integration times for a 6-m HabEx-like configuration.

Table 1 also gives an indication of the relative merits for several candidate observational strategies. Two of them use observations at a single phase angle and the other two combine three phase angles. The selection of phase angles couples to the number of visits required, a facility scheduling issue. We assume each phase angle would require one visit. If scheduling allowed for only one visit, what would be the priority? Several factors play into this question. First, for a circular orbit, star-planet angular separation will be maximum at phase angle 90°, and will be observationally accessible for any system inclination provided that the observation can be timed to coincide with phase angle 90°. On that basis we chose 90° as one value in Table 1. We also examined 60° phase angle which is a compromise between lower phase angles having a higher scattered flux (requiring shorter integration times), but also with a higher probability of falling inside the inner working angle, or not available if the system inclination is less than 30°. Our multiple scattering calculations reveal that reflected flux at 60° phase is about a factor of two higher than reflected flux at 90° phase. In Table 2 we give two examples of minimum and maximum phases limited by system inclination, and minimum star-planet distance limited by the inner working angle (70 mas, from the HabEx study) and by system inclination.

As third and fourth options we examined observations made at three phase angles (60°, 90°, 120°) and (30°, 40°, 50°) to see if the combined phase angle coverage provided significant improvement over a single phase angle. The triple (30°, 40°, 50°) samples the rainbow feature for the larger particles. Karalidi et al. (2011, 2012) called attention to the strong polarization signature in the phase angle range 30° - 40° and recommended sampling in this region to look for clouds of liquid water. Our goal here is to see if measurements can distinguish between water and sulfuric acid with concentrations 25% and 50%.

| Table 2 Planetary System Parameters for Two Cases, and Observationally Accessible Phase Angles | | | | | |
|---|---|---|---|---|---|
| Star | Distance (Parsec) | Star-planet separation (AU at 70 mas) | Inclination (Deg.) | Minimum Phase | Maximum Phase |
| Tau Ceti | 3.7 | 0.26 | 35 | 55 | 125 |
| Fomalhaut | 7.7 | 0.54 | 65.6 | 24.4 | 155.6 |
| Notes for Table 2. Inclination for Tau Ceti from Lawler, et al. 2014. For Fomalhaut from MacGregor, 2017. Scale from Hu et al., (2021). | | | | | |

Table 1 shows that the more complete phase angle coverage does indeed reduce the probability (P1) of mis-identification if minimum $\chi^2$ is the basis for the retrieval. However, the probability of finding an incorrect solution with $\chi^2$ less than 2 (P2) is not significantly reduced. This is true even for the set of three phase angles 30°, 40°, and 50° chosen to sample the rainbow region. The polarization peak in the rainbow region, for particles larger than about 0.25 µm at wavelength 500 nm, is a distinguishing feature that could be used to detect a liquid water cloud (Karalidi et al., (2011 and 2012)) or a sulfuric acid cloud, but the use of the phase angle triplet [30°, 40°, 50°] to distinguishing among them is only a little better than for the triplet [60°, 90°, 120°] and only for the shortwave case. However, Mie calculations show that the rainbow polarization feature shifts a few degrees in phase angle between water and sulfuric acid, so it is conceivable that a phase angle sample tuned to take maximum advantage of that shift, rather than the 10° intervals we investigated, might be more sensitive to composition.

Our study and those of Karalidi et al. (2011 and 2012) differ in other ways as well. Our study is oriented toward a Venus analog, whereas the Karalidi et al. studies are oriented toward an Earth analog. Accordingly, we consider very large (essentially semi-infinite) cloud optical depth with no contribution from Rayleigh scattering by gas above the cloud. Consequently, polarization integrated globally is diluted by multiple scattering more so than for the Karalidi et al. models which, for the most part, have smaller optical depths, with a black surface at the bottom. Planetary reflected flux is another consequence of this difference, with implications for integration times required to achieve the desired signal/noise ratio. Reflected flux will be higher for the semi-infinite cloud than for the low-optical-depth cloud. Karalidi et al. (2011) consider a wide range of optical depth, including some large optical depth cases; Karalidi et al. (2012) examine partial cloud cover with optical depth in cloudy regions generally 10 or less. In the next section we examine integration times applicable to our exo-Venus models, required to make meaningful polarization measurements as per Table 1.

### 3.2 Observational Considerations and Signal/Noise Estimation

#### 3.2.1 Observational Considerations

In Section 3.1 we examined several cases for measurements made with different configurations of wavelength coverage and phase angle sampling. The wavelength passbands (shortwave and longwave) were based on the HabEx study (Gaudi et al. 2020). Although that study considered a 4-m diameter primary mirror and here we consider a 6-m diameter, we assume the same inner working angle (IWA, 70 mas) for the shortwave channel. At the IWA the anticipated contrast is $10^{-10}$. If the longwave measurements are to be made, the distance between the starshade and telescope would need to be reduce with a corresponding change of IWA to a larger value, perhaps to 140 mas.

In consideration of what phase angles go into a measurement set it is important to keep in mind how phase angle folds into the IWA limit and how that maps into the minimum star-planet separation in AU, with implication for flux from the planet. Flux from a directly-imaged planet depends on, among other things, the square of the star-planet distance and the phase angle. A planet observed at phase angle $\theta$ cannot be observed closer to its star than $D\ IWA/\sin\theta$ where $D$ is the distance from the telescope to the star and IWA has units of radians. To exceed the IWA, the minimum star-planet distance for a system at the distance of Fomalhaut is 0.54 AU for an observation at 90° phase; 1.08 AU for an observation at 30° phase. The corresponding numbers for the distance to Tau Ceti are 0.26 AU and 0.52 AU. Scattered flux from the planet is also a function of cloud model parameters and phase angle, decreasing to zero at 180° phase. Smaller cloud optical depth produces greater polarization due to the relative importance of single versus multiple scattering, but the polarized and unpolarized fluxes both decrease as optical depth decreases (as shown in Fig. 7 of Karalidi et al. (2011)), and this works to reduce S/N. Finally, as already mentioned, phase angle limits are coupled to system inclination.

#### 3.2.2 Signal/Noise Estimation

As a point of reference, we leverage the shortwave signal/noise (S/N) calculations made for a notional 4-m HabEx starshade configuration (Hu, et al. 2021). No S/N calculations exist for the longwave spectrum. Noise estimates for the SW are calculated based on what is needed need to subtract backgrounds. They are simulated by SISTER, the Starshade Imaging Simulation Toolkit for Exoplanet Reconnaissance (Hildebrandt et al. 2021). The shortwave detector is a photon-counting EMCCD. Hu et al. (2021) also discussed ways in which noise can be greater than solely from photon counts. For simplicity, we assume noise is due to a small amount of detector noise, but dominated by photon counts from a variety of sources as described next.

We take into account the major noise sources in the starshade-assisted exoplanet imaging, including leaked starlight from imperfect starlight suppression, solar glint by the edge of the starshade, as well as the contribution to the backgrounds from zodiacal and exozodiacal light. The current best estimates of the starshade optical performance are used in the calculations.

We make use of Equation (1) of Hu et al. (2021):

$$S/N = \frac{N_P}{\sqrt{N_P + \alpha(N_S C + N_G + N_E + N_Z + N_D) + \beta^2 (N_S C + N_G + N_E + N_Z + N_D)^2}} \quad (1)$$

Where $N_P$, $N_S$, $N_G$, $N_E$, $N_Z$, and $N_D$ are the counts from the planet, the star, the starshade glint, the exozodiacal dust, the local zodiacal dust, and detector noise, respectively, and C is the instrument contrast. Adapting the code for polarization, we left unchanged values for $N_S C$ ($C$ = 10-10) $N_G$, $N_E$ (for exo-zodi = 20.8 mag/square arc-second), $N_Z$, (for local zodi = 23 mag/square arc-second), and $N_D$ (includes thermal dark counts plus clock-induced charge) but we multiplied all terms except detector noise by the factor 0.5 to simulate imaging through a linear polarizer of an unpolarized source. Although the assumption of unpolarized sources is not strictly valid, a simulation for polarized sources gives essentially the same result when the errors for images in two orthogonal polarizers are propagated. From symmetry considerations the two orthogonal polarizers need to be oriented with principal polarization axes parallel and perpendicular to the scattering plane. The glint contribution ($N_G$) assumed a worst-case position angle of the planet, where the glint is maximized (see Hu et al. (2021)), a pointing that would be avoided, if possible, for a polarization observation.

The paramemter α is included in Eq. 1 to account for imaging subtractions. To detect the planet and measure its flux, subtraction of the background would be typically necessary, and this procedure gives α a value of 2. The parameter β takes into account how well the background sources can be subtracted from the image. This will depend on how many pixels can be used to model each source, and the spatial complexity of the source. For sources that extend over many pixels (such as the zodi, which is constant over the entire image) and an exozodi which can be modeled as a ring with smooth radial and azimuthal variation, β can approach zero (meaning background subtraction to the photon-noise limit). If there is a great deal of structure in the exozodi, the substraction may be imperfect and β may be nonzero. Following Hu et al. (2021), we assume background subtraction to the photon noise limit for the values of the parameters α = 2 and β = 0. For a detailed discussion of these choices see Hu et al. (2021).

We take $N_P$ from a scattering (VLIDORT) result as described above, as a function of wavelength and phase angle. The telescope primary mirror diameter is assumed to be 6 m and the system quantum efficiency is set at 30%, as per the HabEx concept (Gaudi et al. 2020).

The code is designed to calculate Signal/Noise for reflected flux, not polarization. To measure polarization, at least two images are needed, with polarizers oriented parallel and perpendicular to the scattering plane. To simulate this, we multiplied each term (except detector noise) in Equation 1 by 0.5 (the effect of passing unpolarized light through a linear polarizer), then propagated the resulting variance for image subtraction: $v_p = v_1 + v_2$ where $v_p$ is the variance in the subtracted image, $v_1$ and $v_2$ are the variances for each of the orthogonal polarizer images and the errors (uncertainties) are $\sigma = \sqrt{v}$ for each term. We note $v_2 = v_1$ for unpolarized sources. Image subtraction in general gives only one of two Stokes parameters needed for the calculation of P, but symmetry of the reflected light allows us to equate that with polarization. This would not be the case if the principal axes of the polarizers were not parallel and perpendicular to the scattering plane. We assume the desired orientation can be achieved by orienting the telescope or polarizing elements.

The original code (Hu et al. (2021)) assumes a spectral resolution of 140 which, for a nominal wavelength of 700nm, gives a bandwidth of 5 nm for spectral measurement. We modified that to 100 nm to improve S/N for polarization measurement. The results of our S/N calculations are shown in Figure 2.

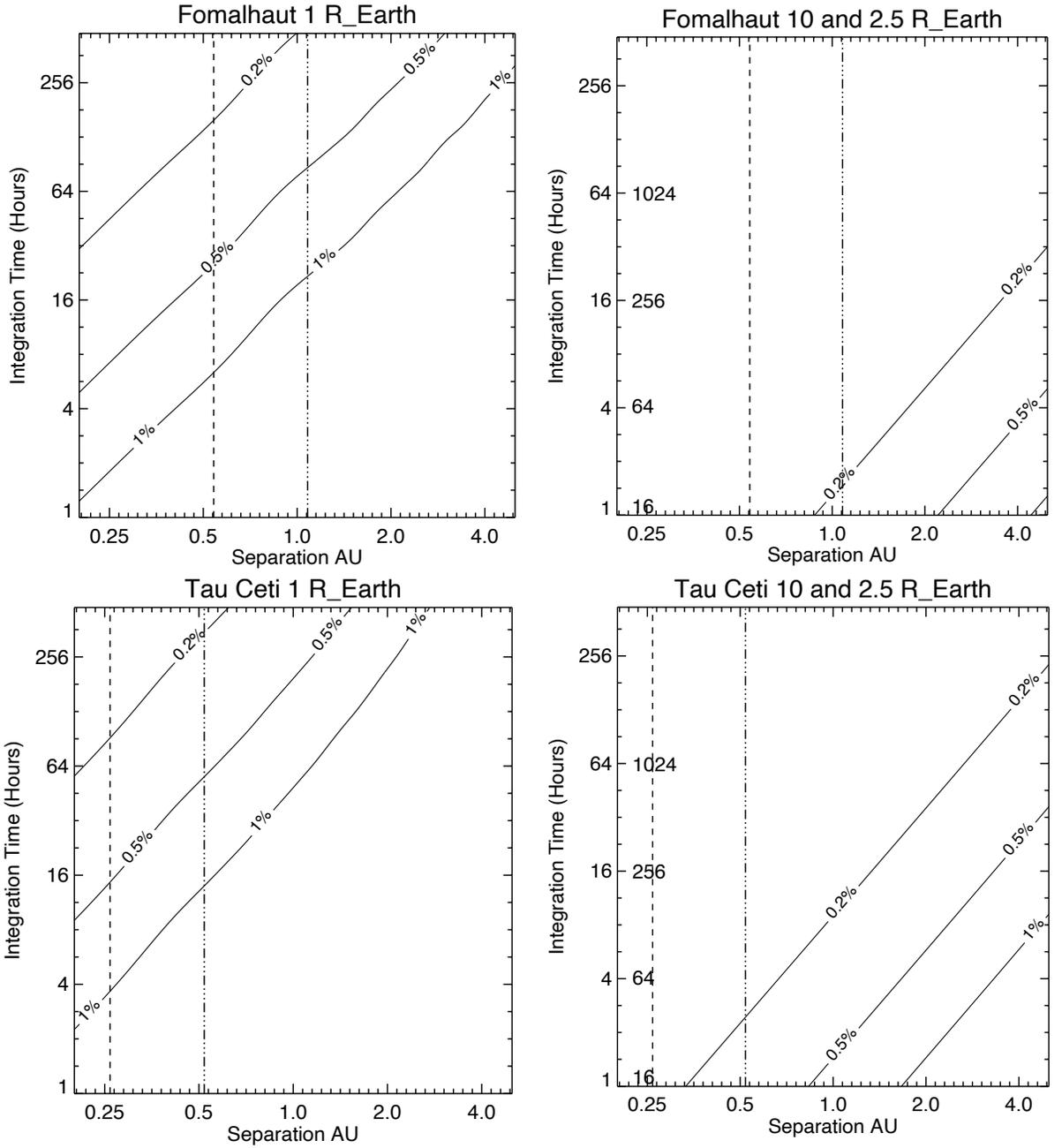

*Figure 2.* These four panels show contours of uncertainty σ for a single passband (0.1 μm, centered on 0.7 μm) for six cases (two example targets and three planet radii as indicated, as functions of star/planet distance in Astronomical Units and integration time ( in the right panels the hours labels inside the borders apply to the case R = 2.5 R_Earth). Contours are shown for three cases (as per Table 1), σ = 1%, 0.5%, and 0.2% of the flux from the planet for a measurement consisting of two images in orthogonal polarizers. The integration time is the sum of integration times for two images in orthogonal polarizers obtained from two sequential exposures. These are for a phase angle of 90° with flux computed from the VLIDORT multiple scattering model. Other parameters are described in the text. The vertical dashed line shows the distance that corresponds to a 70 mas inner working angle, appropriate for phase angle 90°. The dash-dot line corresponds to the IWA limit for a phase angle of 30° (for sampling the rainbow).

## 4. INSTRUMENTAL POLARIZATION

In this section we discuss instrumental polarization for a 4-m HabEx concept based on design records, and look ahead to a possible 6-m space telescope as recommended by the Astro2020 document. Precision measurements of small amounts of polarization that originate in object space are difficult to make. These can be biased because of a large or unknown amount of polarization introduced by the telescope/instrument system. Time consuming and complicated optical instrument system calibration is usually required to deal with this problem. However, for a possible 6-m system, in order to achieve $10^{-10}$ static contrast levels, there is a requirement, derived from the detailed study of the 4-m HabEx, to control instrumentally-induced polarization to a minimum (Davis 2018; Davis et al. 2019; Breckinridge et al. 2019; Krist et al. 2019). These contrast levels cannot be achieved at high system transmittance levels without careful balance of the system to minimize instrument polarization, and this minimum is at the level to assure that object space polarization will be measured to the 1% level.

The 4-m HabEx (Martin et al. 2017) requires a curvature on the primary mirror such that the focal ratio not be less than F 2.5 to maintain surface-ray intercept angles small enough to minimize system polarization effects. A 6-m primary will have the same requirement. The sources of instrument polarization are the highly reflective anisotropic metal thin films, the number of mirror surfaces and the optical configuration or layout. New technology, however is needed to achieve the primary mirror metal thin film polarization reflectivity isotropy to yield both the required 1% surface uniformity in intensity reflectivity and the 10 pm thin film optical depth at mid-spatial frequencies across the entire large surface (either 4 or 6-meter) monolithic primary mirror (Shaklan & Green 2006). Diffraction-limited imaging, necessary for exoplanet coronagraphy, requires that the complex electric field be coherent (Mandel & Wolf 1965) at all points across the entire aperture simultaneously, as viewed from the image plane (Breckinridge 2012).

The 4-m HabEx opto-mechanical layout (Martin et al. 2017, see Figure 9 and Table 7) includes at least one optical subsystem or element that significantly alter the polarization aberrations. One is the planned use of dichroic filters (Heath et al. 2020) to divide wavefronts into separate wavelength bandpasses to isolate optical spectral regions of interest for coronagraphy. The other is the polarization subsystem implemented by the designer to partition the wavefront into two parts to correct independently, two separate polarization aberrations (Martin et al. 2017). Innovative opto-mechanical layout design may eliminate the need for both of these optical subsystems to create a more polarization-neutral system. A 6-meter design would be a big enough change that a different optical path will be required than that used for the 4-meter HabEx and its design should have a requirement to be as polarization neutral as possible and minimize the number of optical surfaces available to scatter and polarize light (Breckinridge & Lillie 2016).

To conclude, it should be possible to design a polarization-neutral telescope and instrument to enable polarization measurements at the 0.2% level or better. The design specification for instrumental polarization for the Daniel K Inouye Solar Telescope is less than five parts in $10^4$ (Rimmele et al. 2020). To achieve the desired limit on instrument polarization the science community would need to levy requirements on the optical engineering community to ensure this will happen.

## 5. DISCUSSION

Section 3 described a framework in which to consider feasibility of using polarization measurements to assess sulfuric acid concentration and habitability for clouds on an exoplanet in the habitable zone. Such an assessment requires a realistic evaluation of the uncertainties. For these we used a software package (Hu et al. (2021)) based on an instrument model (Hildebrandt et al. 2021), for a notional HabEx mission with a 4-m diameter telescope primary (Gaudi et al. 2020), but extended to 6-m. Our findings are applicable to a 6-m diameter telescope as recommended by the Astro2020 document.

Uncertainty contours are shown in Fig. 2, for a single measurement at 0.7 µm, with passband 0.1 µm, and phase angle 90°. This, combined with probabilities shown in table 1, makes it possible to judge what set of observations and what integration times would be needed to retrieve sulfuric acid concentration in the range 0% to 50%. Within that range, the possibility of life as we know it, in cloud droplets, ranges from possible to impossible (Hallsworth et al. 2021). Specifically, for a planet with a radius equal to that of the Earth, integration times must be longer than ~8 hours for the Fomalhaut case, for a planet just outside the inner working angle (70 mas). For Tau Ceti, the inner working angle maps to a smaller planet-star distance, resulting in a shorter minimum integration time (~ 4 hours). To measure polarization at the 1% level for planets at greater orbital/angular distances would require longer integration times. Planets with larger radii can be sensed with shorter integration times (right panels in Fig. 2).

The calculations employed nominal values for the parameters that affect s/n. Exozodi was taken to be 3 zodies (1 zodi being the intensity of the solar system zodiacal light viewed from outside the solar system). The true value for any given system in the target list will be different, but the value we use is in the likely range (Ertel et al. 2020), and is the same as the starting value for the study by Hu et al. (2021). Other parameters listed in Section 3.2 have traceability in terms of laboratory and numerical investigation. Of particular interest is the glint, which depends on the angle of sunlight reflected toward the telescope. We use a worst-case value, where the angle is in the specular regime.

Polarization measurements can lead to determination of the real part of the refractive index of the cloud droplets, and sulfuric acid concentration can be inferred from that. If the refractive index differs from that of water there remains an ambiguity in the mapping of refractive index to composition (e.g. a salt solution as a component along with sulfuric acid – see Mogul et al. 2021; Rimmer et al., 2021; Bains et al., 2021). To better secure the compositional assignment for an exoplanet it may be possible to detect gas-phase components such as sulfur dioxide spectroscopically. Sulfuric acid spectral features exist at near-infrared wavelengths between 1 and 1.8 µm, as well as at mid-infrared wavelengths ~7 µm (e.g., Hu et al. (2013)). The baseline maximum wavelength for Habex imaging at 10-10 contrast is 1 µm. At longer wavelengths the contrast degrades. However, for imaging at 1.8 µm, a 10-10 contrast would be possible if the distance between the telescope and starshade is shortened relative to the shortwave baseline. The penalty would be a larger inner working angle which would reduce the regions where S/N is highest.

Discussion thus far has dealt with the possibility of determining sulfuric acid concentration in cloud droplets as Hansen and Hovenier did for Venus. Although we cannot sense the surface for a planet covered by optically thick clouds, in what follows we comment on possible implications for surface geochemistry and habitability, provided that surface temperature is low enough to support liquid water/sulfuric acid. This is not the case for Venus, and so our discussion next departs from a strict Venus analog, but may be a viable scenario for cloudy exoplanets that either have a less effective greenhouse (or even an anti-greenhouse - Tabor et al. 2020) or receive less irradiance than does Venus.

Detection of sulfuric acid in clouds can have implications for geochemistry and, secondarily, for habitability in lakes. Acidic clouds facilitate chemical weathering of land surface. The weathering reactions produce dissolved ions that are carried into bodies of water to increase salinity and reduce water activities in lakes and oceans. Increasing salinity decreases water activity because dissolution of non-volatile solutes lowers the solution's equilibrium vapor pressure. For lakes, this can drive water activity to prohibitive levels for life. For example, typical water activity in the Dead Sea is 0.69 (Krumgalz et al. 2000), approaching the currently known lower limit of 0.585 for life (Hallsworth et al. 2021, Stevenson et al. 2017). Krumgalz et al. (2000) estimate that evaporation under current climate conditions can eventually reduce Dead Sea's water activity to 0.50, well below the limit. For oceans, reaching such low water-activity levels is very unlikely; for reference, Earth's seawater's activity is ~ 0.98 (Benison et al. 2021). Nonetheless, we note that metabolism ceases for the vast majority of microbial species below 0.90 water activity (Grant 2004, Benison et al. 2021). Therefore, while sulfuric acid is not expected to prevent life on an oceanic scale, it can drastically constrain the diversity of life, and thus, range of plausible biosignatures.

In a broader context, sulfur, with valence states ranging from -2 to +6, has been the dominant oceanic redox buffer for the past 1.8 billion years on Earth and played a controlling role in the upper ocean's redox state before 3.2 Ga (Huston & Logan 2004). Prior the Great Oxidation Event (2.4 Ga) sulfur was mostly in the reduced, insoluble form of pyrite, so dissolved Fe(II) was more dominant in controlling the bulk ocean redox state. Redox chemistry is the fundamental energy source for life, and redox state is a key parameter in determining the chemical speciation of all elements, and thus nutrient availability, in the ocean. Aside from being a redox buffer, sulfate is a prevalent energy source for anaerobic metabolism. Indeed, evolution of life and planetary redox state on Earth has been intricately coupled with sulfur chemistry and the global sulfur cycle, staring with one of the oldest lifeforms on Earth, sulfate-reducing microorganisms (Huston & Logan 2004, Rollinson 2007). Such evolution is manifest in different predicted biosignatures for exoplanet observations (e.g. see Figure 3-2 in the LUVOIR Mission Concept Final Report, https://asd.gsfc.nasa.gov/luvoir/reports/). Hence, detection of sulfuric acid in clouds, which indicates a global presence of soluble sulfur, will provide a fundamental piece of information regarding ocean redox state (if an ocean exists) and biosignatures on the observed planet.

## ACKNOWLEDGMENTS

This research was carried out at the Jet Propulsion Laboratory, California Institute of Technology, under a contract with the National Aeronautics and Space Administration (80NM0018D0004) and funded through JPL's internal Research and Technology Development program. Part of the research was supported by the NASA Exoplanet Research Program grant # 80NM0018F0612. The idea for this work sprang from a workshop sponsored by and held at the Lorentz Center, Leiden, The Netherlands, in 2019.